\documentclass[sigconf, nonacm]{acmart}
\acmConference[EASE 2025]{The 29th International Conference on Evaluation and Assessment in Software Engineering}{17–20 June, 2025}{Istanbul, Turkey}

\usepackage{amsmath,amsfonts}
\usepackage{algorithmic}
\usepackage{graphicx}
\usepackage{textcomp}
\usepackage{xcolor}
\usepackage{soul}
\usepackage{epstopdf}
\usepackage{booktabs}
\usepackage{multirow}
\usepackage{url}
\usepackage{flushend}

\graphicspath{{./images/}}

\begin{document}

\title{Towards an Interpretable Analysis for Estimating the Resolution Time of Software Issues}

\author{Dimitrios-Nikitas Nastos}
\orcid{https://orcid.org/0009-0007-2240-2835}
\affiliation{%
\institution{Electrical and Computer Engineering Dept., Aristotle University of Thessaloniki}
\city{Thessaloniki}
\country{Greece}
}
\email{diminast@ece.auth.gr}

\author{Themistoklis Diamantopoulos}
\orcid{https://orcid.org/0000-0002-0520-7225}
\affiliation{%
\institution{Electrical and Computer Engineering Dept., Aristotle University of Thessaloniki}
\city{Thessaloniki}
\country{Greece}
}
\email{thdiaman@issel.ee.auth.gr}

\author{Davide Tosi}
\orcid{https://orcid.org/0000-0003-3815-2512}
\affiliation{%
\institution{Department of Theoretical and Applied Sciences, Universit\`a degli Studi dell’Insubria}
\city{Varese}
\country{Italy}
}
\email{davide.tosi@uninsubria.it}

\author{Martina Tropeano}
\affiliation{%
\institution{Department of Theoretical and Applied Sciences, Universit\`a degli Studi dell’Insubria}
\city{Varese}
\country{Italy}
}
\email{mtropeano@studenti.uninsubria.it}

\author{Andreas L. Symeonidis}
\orcid{https://orcid.org/0000-0003-0235-6046}
\affiliation{%
\institution{Electrical and Computer Engineering Dept., Aristotle University of Thessaloniki}
\city{Thessaloniki}
\country{Greece}
}
\email{symeonid@ece.auth.gr}

\begin{abstract}
Lately, software development has become a predominantly online process, as more teams host and monitor their projects remotely. Sophisticated approaches employ issue tracking systems like Jira, predicting the time required to resolve issues and effectively assigning and prioritizing project tasks. Several methods have been developed to address this challenge, widely known as bug-fix time prediction, yet they exhibit significant limitations. Most consider only textual issue data and/or use techniques that overlook the semantics and metadata of issues (e.g., priority or assignee expertise). Many also fail to distinguish actual development effort from administrative delays, including assignment and review phases, leading to estimates that do not reflect the true effort needed. In this work, we build an issue monitoring system that extracts the actual effort required to fix issues on a per-project basis. Our approach employs topic modeling to capture issue semantics and leverages metadata (components, labels, priority, issue type, assignees) for interpretable resolution time analysis. Final predictions are generated by an aggregated model, enabling contributors to make informed decisions. Evaluation across multiple projects shows the system can effectively estimate resolution time and provide valuable insights.
\end{abstract}




\maketitle

\section{Introduction}
Nowadays, developers host their projects online in code hosting facilities and manage them with issue tracking systems like Bugzilla or Jira. These services enable developers to co-manage projects, implement new features, fix issues, plan sprints, craft product releases and generally keep track of all the management aspects of their project. This collaborative paradigm has also brought forth an abundance of data, which can be used to automate certain tasks with the goal of reducing the time and effort required by the developers.

Indeed, several researchers have sought to confront challenges related to issue monitoring and assignment, including e.g.\ the problem of finding the most suitable developer for fixing an issue \cite{MurphyCubranic:BugTriaging1, Anvik:BugTriaging2, Matsoukas:IssueAssignment, Alkhazi:Commits} or determining the priority/severity of an issue \cite{Sharma:Priority1, Tian:Priority2, KanwalMaqbool:Priority3, Diamantopoulos:TaskImportancePrediction, Lamkanfi:CoarsegrainedSeverity1, Yang:CoarsegrainedSeverity3}. These efforts are often coupled with the challenge of \textit{bug-fix time prediction}, i.e.\ the problem of determining how much time is needed in order to fix an issue. Practically, one could even argue that all these problems are connected, considering that knowing the effort required to resolve an issue is necessary for determining the most suitable assignee or even for prioritizing it \cite{Giger:BugFixTime}.

In fact, several approaches have lately been proposed for the challenge of issue resolution time prediction
%In fact, the problem of issue resolution time prediction has lately attracted the attention of the research community and current approaches employ various methods to predict it 
\cite{Weiss:BugFixTime, Panjer:BugFixTime, Giger:BugFixTime, BhattacharyaNeamtiu:BugFixTime, Marks:BugFixTime, LamkanfiDemeyer:BugFixTime, Zhang:BugFixTime, Habayeb:BugFixTime}. 
Certain methods are based on textual data and machine learning models \cite{Weiss:BugFixTime, Panjer:BugFixTime}, while others employ also information on the contributors/components involved \cite{Giger:BugFixTime, BhattacharyaNeamtiu:BugFixTime, Marks:BugFixTime, LamkanfiDemeyer:BugFixTime}. There are also several interesting approaches using Markov models \cite{Zhang:BugFixTime, Habayeb:BugFixTime} or even deep learning \cite{Sepahvand:BugFixTime, Lee:BugFixTime}. 

Although efficient, most of these approaches suffer from important limitations. 
Most of them focus 
%The majority of works perform an analysis based 
on past issue data taking into account only the text (titles, descriptions), disregarding the metadata of issues, such as their priority or labels. Furthermore, they do not always employ semantics to unveil correlations among the issues and the components of the project. %What is more important, however, is that 
Moreover,
the estimates %provided 
are black-box, i.e.\ they are not interpretable under the prism of expected effort, expertise and priority. 
The
%For instance, they typically estimate the 
time required to resolve an issue
is often estimated without taking into account its severity (i.e.\ prioritize critical issues %are prioritized 
against others), whether the issue is hard to fix (and thus it requires more effort), or if the developer undertaking it is experienced in the relevant component/topic (and thus it may demand less/more of his/her time).
%Especially for bug-fix time approaches, which are the focus of this paper, most approaches view fix time purely as the effort required to resolve an issue and disregard other dimensions that may influence fix time, such as e.g.\ the priority of the issue or even the expertise of the developer that is assigned to it.

%In this work, we present an issue monitoring approach that helps contributors better understand the specifics of each issue, in order to determine the actual effort required to resolve it in accordance with its severity, and thus prioritize it accordingly. We perform our analysis on a large dataset of Jira issues using the BERTopic topic modeling technique \cite{Grootendorst:BERTopic} to extract the semantics of issues. Furthermore, we take into account the metadata of issues to build an interpretable model that can support the decisions of the triager providing not only a probability value but also, most importantly, a comprehensive response about the estimated fix time of an issue.

In this work, we present an interpretable issue monitoring approach that goes beyond black-box predictions by leveraging both textual and metadata-driven insights. Unlike prior methods that rely primarily on textual data, our approach incorporates issue priority, component, and assignee history to provide a structured and explainable resolution time estimate. By using BERTopic \cite{Grootendorst:BERTopic} for semantic analysis and integrating historical metadata% trends
, we enable informed decision-making for issue %triaging 
and sprint planning.

The rest of this paper is organized as follows. Section \ref{sec:relatedwork} reviews the related work in the challenge of issue resolution time prediction and further discusses the limitations of existing approaches. Our methodology for building a new issue resolution time assessing mechanism is presented in Section \ref{sec:methodology}. Section \ref{sec:evaluation} evaluates our approach using a set of software projects and illustrates its interpretability. The implications of our methodology and the relevant insights %in the area of resolution time estimation 
are discussed in Section \ref{sec:discussion}. Finally, Section \ref{sec:conclusion} concludes this work and provides ideas for future work.

\section{Related Work} \label{sec:relatedwork}
Issue resolution time prediction has emerged as a key research challenge in software engineering, attracting a variety of approaches. Giger et al. \cite{Giger:BugFixTime} used decision tree analysis to predict whether issues would be fixed quickly or slowly, relying on both static issue attributes and early post‐submission changes; while their models outperform random classification, the binary fast/slow categorization and moderate interpretability leave finer-grained questions unanswered. Weiss et al. \cite{Weiss:BugFixTime} adopted text mining techniques to match new issue reports with similar past reports for effort estimation, yet their reliance on report text quality limits insight into how different features contribute to predictions. Panjer \cite{Panjer:BugFixTime} investigated Eclipse bug lifetimes using various data mining techniques and basic attributes such as cc lists and comments, but his single-project focus and lack of broader interpretability limit its applicability. Bhattacharya and Neamtiu \cite{BhattacharyaNeamtiu:BugFixTime} compared different bug-fix time models and demonstrated incremental improvements in prediction accuracy; however, their work does not offer a holistic explanation of critical factors influencing fix time. 

Marks et al.~\cite{Marks:BugFixTime} studied resolution times across %large
open source projects using a random forest classifier to capture %complex
interactions among fixing time, component information, and other features, yet the ensemble nature of their approach obscures the underlying decision logic. Lamkanfi and Demeyer \cite{LamkanfiDemeyer:BugFixTime} focused on filtering bug reports to enhance fix-time analysis, reducing noise but yielding models that remain opaque regarding feature contributions. Zhang et al. \cite{Zhang:BugFixTime} conducted an empirical study on commercial projects using a variety of project-specific features; although their results validate industrial applicability, the complexity and tuning of their models hinder interpretability and generalizability. Habayeb et al.~\cite{Habayeb:BugFixTime} employed a hidden Markov model to capture latent dynamics in issue resolution, effectively modeling sequential dependencies but operating as a black box with limited transparency regarding observable influences. Sepahvand et al.~\cite{Sepahvand:BugFixTime} applied word embedding and deep long short-term memory networks to predict resolution time, capturing complex textual and structured data patterns at the expense of transparency, while Lee et al.~\cite{Lee:BugFixTime} introduced a continual prediction framework using deep learning-based activity stream embedding that adapts over time but suffers from similar interpretability issues. Qiao et al.~\cite{Qiao2024PredictingIR} proposed a refined prediction model using static and dynamic features from GitHub projects, improving accuracy but still facing issues with generalizability and interpretability. Özkan et al.~\cite{zkan2024BugAT} developed a domain-specific estimation approach for network software projects using structured Jira data, which performs well in local use cases but lacks broad applicability.

% TODO: Write the limitations of current approaches, check fourth paragraph of introduction

Further advancing previous work, our approach leverages robust prediction models that integrate both static and dynamic post-submission data while incorporating interpretable structures, such as rule-based decompositions. This way, we are able to clearly illustrate the relative impact of key features, thereby providing accurate predictions alongside actionable insights for issue triage and resource allocation.

\section{Methodology} \label{sec:methodology}
\begin{figure*}[ht]
\centering
\includegraphics[width=\linewidth]{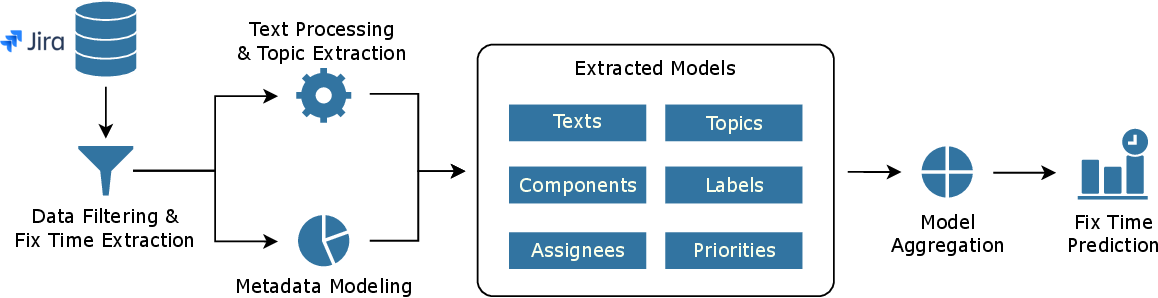}
\caption{Overview of our Issue Monitoring Methodology}
\label{fig:system}
\Description{Overview of our Issue Monitoring Methodology, including text processing, models for all elements (texts, topics, metadata) and model aggregation}
\end{figure*}

%Our methodology aspires to build an interpretable model for predicting issue resolution times while providing meaningful insights on the factors influencing these predictions. 
%To design this system, we use as input a set of Jira projects,
To design our methodology, we utilize a set of Jira projects drawn from the dataset created by Diamantopoulos et al.~\cite{Diamantopoulos:JiraDataset}, which includes over one million issues across 656 Apache projects. As illustrated in Figure~\ref{fig:system}, our approach follows a structured pipeline consisting of four main stages: (1) Data Preprocessing, (2) Issue Resolution Time Extraction, (3) Feature Modeling from Texts and Metadata, and (4) Resolution Time Estimation through Model Aggregation. The full dataset and source code required to reproduce our results are publicly available at: \url{https://github.com/AuthEceSoftEng/issues-fix-time}.

% TODO: Check these out, they may help us
%Practically, we could/should answer the following questions:
%
%\begin{itemize}
%\item If we define the bug lifecycle time as the total time between finding and resolving a bug, how much time is dedicated in actual fixing? And how much time is waiting time?
%\item Is the actual fixing time related to developer expertise?
%\item Is the actual fixing time related to the component/topic/label of the project?
%\item Is the waiting time related to the severity/priority of the issue?
%\end{itemize}

\subsection{Data Preprocessing}
Our analysis is performed on a per-project basis, recognizing that each project has unique characteristics, workflows, and team dynamics. This approach allows us to capture patterns within each development team, resulting in more accurate and relevant predictions.

We apply several preprocessing steps to prepare the data. First, we focus exclusively on resolved issues with assigned developers, filtering out open or unassigned issues that lack reliable resolution time information. For each project, we apply filtering criteria to ensure data quality: requiring both component and label information for sufficient context, and setting a threshold of at least 20 issues per assignee to capture meaningful historical resolution patterns.

\subsection{Issue Resolution Time Extraction}
For each project repository, we divide the overall issue lifecycle into three intervals:

\begin{itemize}
  \item \textbf{Before-Work Time:} The period between issue creation and when development work begins (typically marked by a status change to ``In Progress'').
  \item\textbf{During-Work Time:} The active development period when the issue is being resolved (from ``In Progress'' to ``Resolved'').
  \item\textbf{After-Work Time} The period after the resolution is implemented but before the issue is fully closed (from ``Resolved'' to ``Closed'').
\end{itemize}

Our analysis focuses primarily on the \textbf{During-Work Time}, as this represents the actual effort given by developers to resolve the issue. To maintain consistency across projects, we employ a \textbf{fixed time categorization} to classify resolution times into predefined categories. For the categorization% of issues
, we use absolute time thresholds to ensure comparability between different projects and teams.

These thresholds were selected based on empirical distributions of resolution times across multiple projects and were designed to reflect meaningful distinctions in real-world development workflows. The \textbf{< 0.5 days} category represents minor or trivial issues typically addressed within a few hours%, often during the same working session
. The \textbf{0.5–2 days} category encompasses short, well-scoped tasks that fit comfortably within one to two working days or agile sub-tasks. The \textbf{2–5 days} category captures moderately complex issues requiring deeper investigation, coordination, or multiple iterations. Finally, the \textbf{> 5 days} category indicates long-running or high-complexity issues involving architectural work, significant debugging, or external dependencies.

%For the specific project, add two figures, one showing the different time intervals (before-work, during-work, after-work) and one showing how we split the during-work time into 3 classes (fast, medium, slow).

\subsection{Extracting Issue Texts and Topics}
To understand the semantic content of issues, we apply comprehensive text preprocessing and topic modeling techniques. For each issue, we extract and clean the summary and description fields by removing HTML tags, special characters, numbers, and stopwords. We also apply lemmatization to reduce words to their base forms, enabling more effective pattern identification.
We employ BERTopic \cite{Grootendorst:BERTopic}, a state-of-the-art neural topic modeling technique, to extract meaningful topics from the preprocessed issue texts. BERTopic leverages BERT embeddings to create semantically rich document representations, then applies dimensionality reduction and clustering to identify coherent topics. This approach significantly outperforms traditional topic modeling techniques like LDA in capturing subtle semantic relationships between issues.
For each project, we extract a variable number of topics based on the dataset characteristics, with each topic represented by its most relevant keywords. These topics often align with specific types of issues or functional areas within the project. By mapping issues to topics, we gain insight into how different issue topics affect resolution times.
For textual features, we also utilize text similarity score with previously resolved issues based on BERT embeddings.

\subsection{Modeling Issue Metadata}
In addition to textual content, we model several key metadata elements that influence resolution times:
\begin{itemize}
\item{Components: 
We normalize component names and extract the primary (first listed) component for each issue. Component modeling helps us understand how different technical areas of a project impact resolution times. Some components may be more complex or require specialized knowledge, leading to longer resolution times.
%We normalize component names to handle variations and create a primary component feature from the first listed component for each issue. Component modeling helps us understand how different technical areas of a project impact resolution times. Some components may be more complex or require specialized knowledge, leading to longer resolution times.
}
\item{Labels: Similarly to components, we normalize labels and extract the primary label for each issue. Labels often indicate issue characteristics like "documentation," "enhancement," or "critical-bug," which can significantly affect resolution time expectations.}
\item{Assignees: Developer expertise %and workload are crucial factors 
is crucial in issue resolution. We model it
%assignee performance 
by analyzing %their
historical patterns, including average resolution times and expertise with specific components or topics. This helps identify when an issue might be resolved more quickly due to assignee expertise% or more slowly due to assignee workload
.}
\item{Priorities: Issue priority (e.g., Critical, Major, Minor, Trivial) provides valuable context for resolution time expectations. Higher priority issues typically receive more immediate attention but may also indicate more complex problems.}
\item{Issue Type: Different issue types (e.g., Bug, Task, Improvement, Epic) have varying resolution times. Bugs often require urgent fixes, while improvements and epics may take longer due to planning and development cycles. Including issue type as a feature helps refine resolution time predictions.}
\end{itemize}
For each metadata element, we encode categorical values where necessary and build an independent predictive model using lightweight classifiers such as decision trees and logistic regression models. These models are trained separately to capture the specific relationship between each metadata factor and the resolution time categories. Their outputs contribute to the final resolution time prediction while also providing interpretable insights about how each feature influences the resolution process.

\subsection{Resolution Time Estimation}
Our approach predicts issue resolution times using a stacked machine learning model that integrates structured metadata, textual features, and historical patterns. Rather than relying on a single black-box predictor, we train multiple specialized base models—each capturing signals from issue type, priority, components, labels, assignee history, topic distributions, and textual similarity features.

Each base model independently identifies predictive patterns within its respective feature set. For metadata features (e.g., priority, assignee history, components), we employ lightweight classifiers such as decision trees, random forests, or logistic regression models. For textual features, we use semantic topic vectors extracted through BERTopic and text similarity scores computed via BERT embeddings, both serving as inputs to separate classifiers. Hyperparameters are kept at default scikit-learn settings to maintain simplicity and focus on interpretability.

The outputs from these base models (i.e., predicted class probabilities) are then combined through a stacking ensemble approach. A logistic regression model serves as the meta-learner, learning to optimally aggregate the outputs of the specialized models based on training data, thus producing the final prediction of the resolution time.

\section{Evaluation} \label{sec:evaluation}
\subsection{Experimental Results} \label{subsec:results}

Our methodology is evaluated on a set of software projects extracted from the Apache Jira installation. The evaluation focuses on predicting issue resolution time using a stacked machine learning approach. The selected projects encompass a diverse range of characteristics, including project size, number of contributors, and complexity, allowing for a robust assessment of model performance. Our evaluation uses six projects from the Apache ecosystem: MESOS (4523 issues), IMPALA (5220 issues), BEAM (4703 issues), CASSANDRA (4820 issues), HIVE (5180 issues), and SPARK (4446  issues), each categorized into four resolution time classes. We used an 80\%/20\% stratified random splits, preserving class distributions, without enforcing temporal order.

\begin{figure}[ht]
    \centering
    \includegraphics[width=\linewidth]{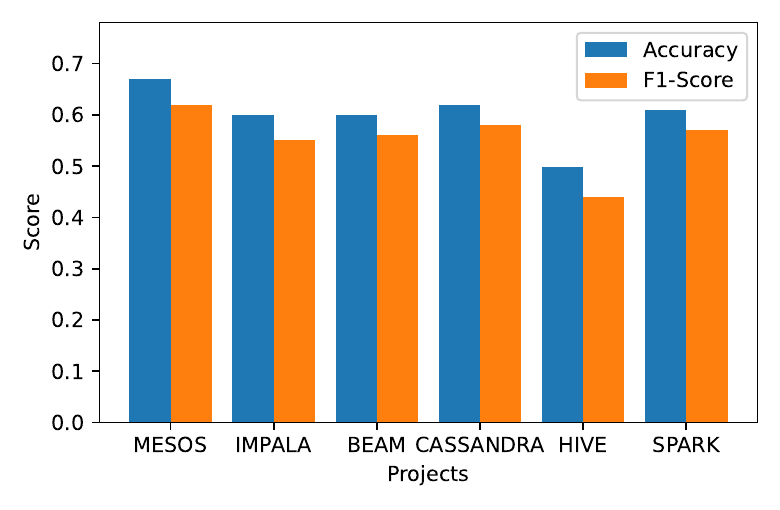}
    \caption{Accuracy and F1-Score of the stacked model across projects.}
    \label{fig:accuracy_graph}
    \Description{Accuracy and F1 of the stacked model across projects.}
\vspace{10pt}
\end{figure}

Given that this is a \textit{multi-class classification problem}, we report both \textbf{accuracy} and \textbf{F1-Score} as evaluation metrics. Accuracy reflects the overall correctness of predictions, while F1-Score balances precision and recall. Figure~\ref{fig:accuracy_graph} presents the accuracy and F1-Score achieved for each project. The model successfully captures meaningful patterns in the prediction of resolution time, with MESOS achieving the highest accuracy (67\%) and F1-Score (62\%), and HIVE the lowest (50\% accuracy, 44\% F1-Score). BEAM and IMPALA achieve moderate performance (around 60\% accuracy and 55\% F1-Score), suggesting that the model generalizes well across different project domains.

Overall, the results demonstrate the applicability of the proposed stacked machine learning approach in predicting resolution time across diverse software projects. The model provides resolution time estimations and establishes a foundation for future enhancements such as workload integration. These findings reinforce the value of automated resolution time prediction in optimizing software development workflows and improving issue resolution strategies.

\subsection{Analysis of the Stacked Model Performance} \label{subsec:casestudy}

To illustrate the performance and reliability of our stacked model, we present a real example output of the system where we demonstrate how the model successfully captures key characteristics of a software issue and assigns the correct resolution time with strong confidence.

\textbf{Test Case: MESOS-7521}
\begin{itemize}
    \item \textbf{Issue Summary:} Major performance regression in DRF sorter
    \item \textbf{Priority:} Blocker
    \item \textbf{Issue Type:} Bug
    \item \textbf{Labels:} ['performance']
    \item \textbf{Components:} ['allocation']
    \item \textbf{Actual Resolution Time Category:} More than 5 days
    \item \textbf{Model Prediction:}
    \begin{itemize}
        \item \textbf{More than 5 days:} 49.0\% \checkmark
        \item 0.5--2 days: 33.0\%
        \item Less than 0.5 days: 13.0\%
        \item 2--5 days: 5.0\%
    \end{itemize}
\end{itemize}

This test case is an indicative example of how the stacked model \textbf{leverages multiple predictive signals} to arrive at a strong and accurate decision. The model confidently predicted \textbf{More than 5 days} due to several contributing factors:

\begin{itemize}
    \item \textbf{Priority-Based Contribution:}
    \begin{itemize}
        \item \textit{Blocker} issues typically require extensive debugging and validation before deployment.
        \item Historical data confirms that Blocker issues frequently exceed 5 days.
    \end{itemize}
    \item \textbf{Component-Based Contribution:}
    \begin{itemize}
        \item The `allocation` component is linked to \textit{deep architectural changes}.
        \item Past issues related to this component have consistently required long resolution times.
    \end{itemize}
    \item \textbf{Label-Based Contribution:}
    \begin{itemize}
        \item The `performance` label strongly correlates with long debugging and optimization cycles.
    \end{itemize}
    \item \textbf{Issue Type-Based Contribution:}
    \begin{itemize}
        \item \textit{Bugs} associated with performance-related regressions often involve complex debugging.
        \item Historical data suggests that performance-related bugs in critical systems tend to require extensive testing before resolution.
    \end{itemize}
    \item \textbf{Assignee-Based Contribution:}
    \begin{itemize}
        \item The assigned developer has a history of resolving \textit{long-duration issues}.
        \item The model learns from this pattern and reinforces the probability of a longer resolution time.
    \end{itemize}
    \item \textbf{Topic and Text-Based Contribution:}
    \begin{itemize}
        \item The model utilizes textual data from issue descriptions to extract recurring patterns in resolution time.
        \item BERTopic clustering reveals that topics linked to architectural changes or optimization tasks align with longer resolution times.
        \item Textual similarity to previous long-duration issues increases the probability of assigning a higher resolution time category.
    \end{itemize}
\end{itemize}

\begin{figure}[ht]
    \centering
    \includegraphics[width=\linewidth]{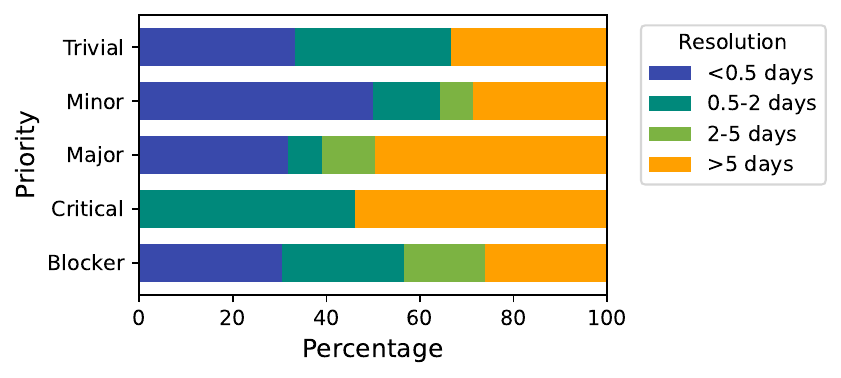}
\vspace{-15pt}
    \caption{Resolution time distribution by priority level.}
    \label{fig:resolution_by_priority}
    \Description{Resolution time distribution by priority level.}
\end{figure}

\begin{figure}[ht]
    \centering
    \includegraphics[width=\linewidth]{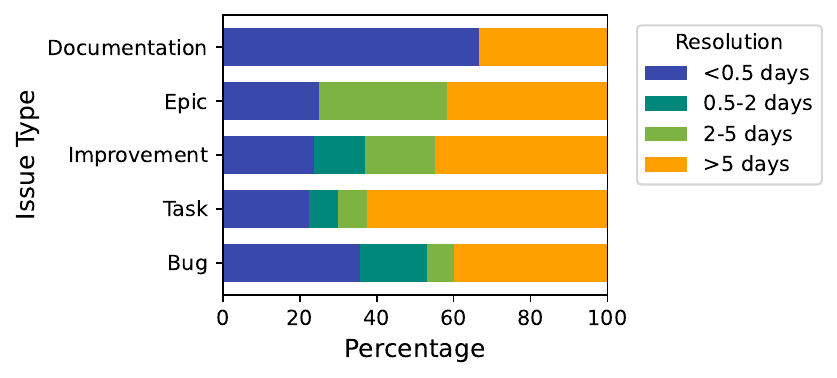}
\vspace{-15pt}
    \caption{Resolution time distribution by issue type.}
    \label{fig:resolution_by_issuetype}
    \Description{Resolution time distribution by issue type.}
\end{figure}

Figures~\ref{fig:resolution_by_priority} and~\ref{fig:resolution_by_issuetype} depict how issue priority and type correlate with resolution times, highlighting the role of metadata.
.This case \textbf{demonstrates the robustness of our stacked model} in capturing \textbf{historical patterns and issue characteristics} to predict resolution time accurately. Unlike ambiguous cases, here the model benefits from \textbf{aligned signals} across priority, component, label, and assignee distributions:

\begin{itemize}
    \item The \textbf{priority} and \textbf{component} factors strongly indicate a longer resolution time.
    \item The \textbf{label-based} feature further supports the prediction.
    \item The \textbf{historical behavior of the assignee} confirms the expected category.
\end{itemize}

As a result, the model \textbf{successfully predicts the correct resolution time with 49\% confidence}, significantly higher than any other category.

This case highlights how our model excels when \textbf{strong patterns exist} in historical data. By incorporating \textbf{priority, component type, labels, and assignee history}, the stacked model effectively generalizes and makes informed predictions with high confidence. This test case showcases the model's ability to deliver reliable results in well-defined scenarios.

\section{Discussion} \label{sec:discussion}
 %Even when there are similar past issue data, the priority of each issue, the expertise of individual developers, even the current workload of each team member all play a significant role on determining the scheduling and the expected time required to resolve each issue.

In this section, we discuss the scope and applicability of the proposed approach in the broader context of bug/issue resolution time prediction, along with open issues for future research. First and foremost, we argue that current research should prioritize providing interpretable analyses of expected resolution times. In contrast to methods optimized for minimizing metrics like mean absolute error, our approach focuses on delivering interpretable predictions that support informed decision-making, offering visibility that black-box models cannot. This aspect is critical for development teams, who must weigh multiple factors—such as issue priority, developer expertise, and workload—when scheduling and estimating issue resolution times.

Our research highlights the vital role of interpretability in issue resolution time estimation. Unlike black-box models, our methodology provides clear, actionable insights by extracting semantic information through topic modeling and presenting results via intuitive visualizations. This enables teams to understand why certain issues take longer, revealing patterns beyond what numerical metrics capture. Most importantly, by distinguishing between total issue lifespan and actual developer effort, our model offers transparency into the real drivers of resolution time, allowing teams to address bottlenecks, optimize workflows, and set realistic expectations.

The aforementioned distinction is actually one of the possible limitations of contemporary approaches, as it is related to the specifics of each project and each development team. As part of agile development, each team may have it own course of action; certain teams may operate in sprints where developers are assigned issues in predefined time intervals, while others may use a more free-form kanban-style approach assigning and setting in progress multiple issues at once (even if they are completed at different times).
To improve on this aspect, we need an approach that is optimized at project level and based on the dynamics of each team, one that would clearly isolate the resolution times of issues and categorize them (e.g.\ as ``Fast'', ``Medium'', or ``Slow'') based on the overall speed of each development team, a variable that may be relevant to the project domain and complexity, and even the team's development practices. Lastly, such considerations would be interesting to view under the prism of human-in-the-loop approaches, given that team intents must be take into account along with the model parameters in order to provide optimal results.

Concerning any threats to the validity of our approach, several factors may influence the variation in model performance across different projects. We hypothesize that resolution time predictability is affected by issue description quality, metadata completeness, team size, and assignee expertise. Projects with well-described issues, consistent metadata labeling, and stable, experienced contributor bases tend to yield more accurate predictions. These observations, while intuitive, warrant more detailed empirical investigation in future work.

Our study also faces certain limitations that may impact the generalization of the results. First, the presence of class imbalance across resolution time categories could influence evaluation metrics, despite the use of stratified sampling. Additionally, the train/test used for the evaluation serves to assess the methods without considering the temporal sequence of the issues (which better reflects what happens in a software project). Moreover, although our system emphasizes interpretability through metadata feature attribution, no direct human evaluation of explanation quality has been conducted.

Finally, although we conceptually compare our feature sets against prior approaches in the literature, direct performance comparisons are limited due to differences in datasets, project sizes, and issue tracking systems. Factors such as the inherent complexity of projects, contributor practices, and metadata usage policies could also affect model performance and generalization. Future work should address these external validity threats by evaluating the system across broader and more diverse industrial datasets.

\balance
\section{Conclusion} \label{sec:conclusion}
%In this paper, we presented an interpretable approach for estimating the resolution time of software issues%, with a particular focus on those managed through Jira
%, through a combination of topic modeling, metadata analysis, and component-based predictions. 
Software development has evolved into a highly collaborative process where predicting issue resolution times is crucial for effective project management. While existing approaches have made progress, they often act as black-box models without explaining the underlying factors influencing resolution times.

Our methodology addresses this limitation by creating a system that predicts resolution time categories while providing interpretable explanations. By analyzing issue texts with BERTopic and modeling metadata elements such as components, labels, assignees, and priorities, we capture semantic and contextual factors affecting resolution times.

Evaluation across multiple software projects demonstrates that our approach effectively categorizes issues into meaningful resolution time classes while offering actionable insights. Project managers and developers can leverage these interpretations to prioritize issues, plan sprints, and identify bottlenecks to improve workflows.

Future work includes accounting for developer workload, enabling cross-project generalization, and conducting human-in-the-loop studies to validate interpretability. We also aim to build a holistic model that jointly addresses issue triaging and resolution time prediction, ultimately enhancing project management and decision-making through transparent analytics.

%%% Future work
%The generalizability of our findings may be limited by the projects we've analyzed. While we've included a diverse set of projects with varying sizes, domains, and development practices, our approach may not perform equally well in all contexts. Projects with unusual workflows, extremely small teams, or highly specialized domains may require adjustments to our methodology.
%Additionally, our analysis is limited to the Jira issue tracking system. Other issue tracking platforms may have different data structures, workflows, or metadata fields that would require adaptation of our approach.
%Despite these threats, we believe our methodology provides valuable insights for most software development teams using Jira, and the core principles could be adapted to similar contexts with appropriate modifications.

\begin{acks}
Parts of this work have been supported by the Horizon Europe project ECO-READY (Grant Agreement No 101084201), funded by the European Union.
\end{acks}

\bibliographystyle{ACM-Reference-Format}
\bibliography{paper}

\end{document}